\documentclass[runningheads]{llncs}
\usepackage[T1]{fontenc}
\usepackage{graphicx}

\usepackage{amsmath}
\usepackage{booktabs}
\usepackage{multirow}
\usepackage{tabularx}
\usepackage{bold-extra}
\usepackage{xcolor, soul}

\newsavebox\CBox
\def\textBF#1{\sbox\CBox{#1}\resizebox{\wd\CBox}{\ht\CBox}{\textbf{#1}}}

\newcolumntype{Y}{>{\centering\arraybackslash}X}
\begin{document}
\title{Keyword Embeddings for Query Suggestion}
%
%
\author{Jorge Gabín\inst{1,2}\orcidID{0000-0002-5494-0765} \and
M. Eduardo Ares\inst{1}\orcidID{0000-0003-3807-5692} \and
Javier Parapar\inst{2}\orcidID{0000-0002-5997-8252}}
%
\authorrunning{J. Gabín et al.}
%
\institute{Linknovate Science, Rúa das Flores 33, Roxos, Santiago de Compostela, A Coruña, 15896, Spain \and
IRLab, CITIC, Computer Science Department, University of A Coruña, A Coruña, 15071, Spain
}
\maketitle              
\begin{abstract}
Nowadays, search engine users commonly rely on query suggestions to improve their initial inputs. Current systems are very good at recommending lexical adaptations or spelling corrections to users' queries. However, they often struggle to suggest semantically related keywords given a user's query. The construction of a detailed query is crucial in some tasks, such as legal retrieval or academic search. In these scenarios, keyword suggestion methods are critical to guide the user during the query formulation. This paper proposes two novel models for the keyword suggestion task trained on scientific literature. Our techniques adapt the architecture of \texttt{Word2Vec} and \texttt{FastText} to generate keyword embeddings by leveraging documents' keyword co-occurrence. Along with these models, we also present a specially tailored negative sampling approach that exploits how keywords appear in academic publications. 
We devise a ranking-based evaluation methodology following both known-item and ad-hoc search scenarios. Finally, we evaluate our proposals against the state-of-the-art word and sentence embedding models showing considerable improvements over the baselines for the tasks.

\end{abstract}

\section{Introduction}
The use of word embeddings~\cite{word_emb_survey} has improved the results of many Natural Language Processing (NLP) tasks, such as name entity recognition, speech processing, part-of-speech tagging, semantic role labelling, chunking, and syntactic parsing, among others. These techniques represent words as dense real-valued vectors which preserve semantic and syntactic similarities between words. More recently, the so-called document and sentence embedding models allow the computing of embeddings for larger pieces of text directly (instead of doing it word by word), getting state-of-the-art results on different tasks.

Nowadays, search engines are outstanding in recommending lexical adaptations or corrections to users' queries~\cite{query_spelling1,query_spelling2}. However, there is still room for improvement in suggesting semantically similar phrases to users' inputs. The few systems that address this task do it by simply using existing word or sentence embedding models, which leads to poor keyword suggestions. 

Keyword suggestion, which consists in recommending keywords similar to the user's input,  is critical in search scenarios where the completeness of the query clauses may dramatically affect the recall, such as academic or legal search~\cite{academic_search,legal_search}. An incomplete query may end up in a \emph{null search session} \cite{null_queries}, that is, the system presents an empty result list to the user. In tasks where use cases are recall-oriented, having an incomplete or even empty results set greatly diminishes the search experience. 

Having good keyword suggestion models in a keyword-based search engine has many benefits. (i) It will help to identify the query intent~\cite{query_intent}, as users will be able to refine their search by adding new query clauses that may be clarifying. (ii) It will promote serendipity~\cite{amplifying_serendipity,discovery_serendipity}, as users may see a recommended keyword that they were not considering but perfectly fits their search interest. (iii) It will help prevent null sessions or incomplete results lists. (iv) Systems may use semantically similar keywords to the user input to perform query expansion~\cite{query_expansion_survey,russell2021interactive} without further user interaction.

In this paper, we leverage \texttt{Word2Vec}~\cite{mikolov2013efficient} and \texttt{FastText}~\cite{bojanowski2017enriching} models' architecture to generate keyword embeddings instead of word embeddings, this meaning that embeddings represent sequences of words rather than a single word, similar to sentence embeddings. Unlike the base models, which use bag-of-words to build the representations, our approaches are based on bag-of-keywords. Thus, the proposed models exploit the annotated keywords' co-occurrence in the scientific literature for learning the dense keyword representations.

Our main aim for producing keyword embeddings is to represent concepts or topics instead of representing just words' semantics or even their contextual meaning like word embedding models do. The keywords' conceptual or topical semantics rely on an explicit human annotation process. First, the annotator will only select proper or commonly used descriptors for the presented documents. In this way, we may assume that those keywords are a good proxy when searching for the documents' topics. Second, authors use a limited number of keywords when annotating documents, so the strength of the semantic relationships among them is higher than the traditional approach of word co-occurrence in windows of free text.

Along with these models, we propose a new method to perform negative sampling, which leverages connected components to select the negative inputs used in the training phase. This method uses the keywords co-occurrence graph to extract the connected components and then select inputs that are not in the same connected component as negative samples.

To evaluate our models' performance, we compare them against a set of baselines composed of state-of-the-art word and sentence embedding models. We both trained the baselines in the keyword data and used the pre-trained models. In particular, we further trained \texttt{Word2Vec}~\cite{mikolov2013efficient} and \texttt{Sentence-BERT}~\cite{reimers2019sentence} (\texttt{SBERT}) on the task data, and we used \texttt{SBERT}, \texttt{SciBERT}~\cite{beltagy2019scibert} and \texttt{FastText}~\cite{bojanowski2017enriching} base models. We carry out the evaluation in Inspec~\cite{hulth2003improved} and KP20k~\cite{meng2017deep}, two classical keyword extraction datasets compound of scientific publications which count with a set of human-annotated keywords (for now on, we refer to keywords annotated by documents' authors or professional annotators as annotated keywords). The results show significant improvements for the task over state-of-the-art word and sentence embedding models. 

The main contributions of our work can be summarized as follows:
\begin{itemize}
    \item \texttt{Keywords2Vec}, a keyword embedding model based on \texttt{Word2Vec}.
    \item \texttt{FastKeywords}, a keyword embedding model based on \texttt{FastText}.
    \item A new method for negative sampling based on connected components.
    \item Baselines and evaluation methods for the similar keyword suggestion task.
\end{itemize}

\section{Related Work}
This section briefly overviews the existing word and sentence embedding models, particularly those used as baselines in the evaluation stage, along with some previous work related to the keyword suggestion task. 

Dense vector representations, a.k.a. embeddings, have an appealing, intuitive interpretation and can be the subject of valuable operations (e.g. addition, subtraction, distance measures, etc.). Because of those features, embeddings have massively replaced traditional representations in most Machine Learning algorithms and strategies. Many word embedding models, such as \texttt{Word2Vec}~\cite{mikolov2013efficient}, \texttt{GloVe}~\cite{pennington2014glove} or \texttt{FastText}~\cite{bojanowski2017enriching}, have been integrated into widely used toolkits, resulting in even more precise and faster word representations.

\texttt{Word2Vec}~\cite{mikolov2013efficient} was one of the first widely used neural network-based techniques for word embeddings. These representations preserve semantic links between words and their contexts by using the surrounding words to the target one. The authors proposed two methods for computing word embeddings~\cite{mikolov2013efficient}: skip-gram (SG), which predicts context words given a target word, and continuous bag-of-words (CBOW), which predicts a target word using a bag-of-words context.

\texttt{FastText}~\cite{bojanowski2017enriching} is a \texttt{Word2Vec} add-on that treats each word as a collection of character \textit{n}-grams. \texttt{FastText} can estimate unusual and out-of-vocabulary words thanks to the sub-word representation. In \cite{joulin2016bag}, authors employed \texttt{FastText} word representation in conjunction with strategies such as bag of \textit{n}-gram characteristics and demonstrated that \texttt{FastText} outperformed deep learning approaches while being faster.

Sentence embeddings surged as a natural progression of the word embedding problem. Significant progress has been made in sentence embeddings in recent years, particularly in developing universal sentence encoders capable of producing good results in a wide range of downstream applications.

\texttt{Sentence-BERT}~\cite{reimers2019sentence} is one of the most popular sentence embedding models and state-of-the-art on the sentence representation task. It is a modification of the \texttt{BERT}~\cite{devlin2018bert} network using siamese and triplet networks that can derive semantically meaningful sentence embeddings.

There are other sentence embedding models based on pre-trained language models. An example of those language models which achieves excellent results when working with scientific data is \texttt{SciBERT}. \texttt{SciBERT}~\cite{beltagy2019scibert} is an adaptation of \texttt{BERT}~\cite{devlin2018bert} to address the lack of high-quality, large-scale labelled scientific data. This model leverages \texttt{BERT} unsupervised pre-training capabilities to further train the model on a large multi-domain corpus of scientific publications, producing significant improvements in downstream scientific NLP tasks.

Even though word and sentence embeddings have been widely studied, the work on keyword embeddings is still limited. Researchers have employed them mainly on tasks like keyword extraction, phrase similarity or paraphrase identification. Several approaches for the keyword extraction task, like \texttt{EmbedRank}~\cite{embedrank}, directly rely on pre-trained sentence embedding models to rank the extracted keywords. However, other approaches such as \texttt{Key2Vec}~\cite{mahata2018key2vec} train their own model for the phrase embedding generation. In particular, the authors propose directly training multi-word phrase embeddings using \texttt{FastText} instead of a classic approach that learns a model for unigram words combining the words' dense vectors to build multi-word embeddings later.

Yin and Sch{\"{u}}tze~\cite{emb_gen_phrases} presented an embedding model for generalized phrases to address the paraphrase identification task. This approach aims to train the \texttt{Word2Vec} SG model without any modification to learn phrase embeddings. They pre-process the corpus by reformatting the sentences with the continuity information of phrases. The final collection contains two-word phrases whose parts may occur next to each other (continuous) or separated from each other (discontinuous).

The phrase semantic similarity task is akin to the one we address in this paper. Many sentence embedding models, including \texttt{SBERT}, are pre-trained in that downstream task. In~\cite{comp_models}, the authors present a composition model for building phrase embeddings with semantic similarity in mind. This model, named Feature-rich Compositional Transformation (\texttt{FCT}), learns transformations for composing phrase embeddings from the component words based on extracted features from a phrase.

Finally, the keyword suggestion task is also a sub-type of query suggestion where all the input and output queries are represented as keywords. Previous works in query term suggestion did not leverage the power of keyword co-occurrence to recommend new terms. Instead, existing query suggestion systems usually approach this task by suggesting terms extracted from the document's content without relying on the relations between these terms. For example, in~\cite {query_term_sugg}, authors propose indexing a set of terms extracted from the corpus to rank them using a language model given an input query. The problem with these approaches is that they depend on the appearance of semantically related terms in the analyzed documents. 

\section{Proposal}
\label{sec:proposal}
In this section, we present two novel keyword embedding models that leverage \texttt{Word2Vec}~\cite{mikolov2013efficient} and \texttt{FastText}~\cite{bojanowski2017enriching} architectures to produce keyword embeddings. We name these models \texttt{Keywords2Vec} and \texttt{FastKeywords}, respectively. Unlike \texttt{Word2Vec} and \texttt{FastText}, these models are not trained on rolling windows over the documents' full text; instead, we only use combinations from the documents' set of keywords as inputs.

The first of them, \texttt{Keywords2Vec}, modifies \texttt{Word2Vec} CBOW architecture to represent each keyword as one item. That is, we use keywords as token inputs instead of words. Additionally, we change how to perform the training of the models; we will explain it later for both proposals. We also evaluated the SG counterpart, but it performed considerably worse than the CBOW, so we do not report them here for brevity.

The second one, \texttt{FastKeywords}, adapts \texttt{FastText} CBOW variant in a more complex way. First, instead of working with words as the bigger information unit, it works with keywords. Second, it always selects each word of the keyword and the keyword itself as inputs, and then, during the \textit{n}-grams selection process, it generates each word's \textit{n}-grams.

Taking the keyword ``\textit{search engine}'' and $\textit{n} = 3$ as an example, it will be represented by the following \textit{n}-grams:
\begin{equation*}
\fbox{\strut sea}\:\fbox{\strut ear}\:\fbox{\strut arc}\:\fbox{\strut rch}\:\fbox{\strut eng}\:\fbox{\strut ngi}\:\fbox{\strut gin}\:\fbox{\strut ine}
\end{equation*}
the special sequences for words:
\begin{equation*}
\fbox{\strut search}\:\fbox{\strut engine}
\end{equation*}
and the special sequence for the whole keyword: 
\begin{equation*}
\fbox{\strut search\:engine}
\end{equation*}

The model uses special sequences to capture the semantic meaning of both words and keywords. Finally, we implemented a weighting system to ignore ``fill'' \textit{n}-grams used when a keyword does not have enough \textit{n}-grams to fill the desired input size. The inclusion of this kind of \textit{n}-grams is needed because the model requires the same input length on every iteration.

Another novel contribution of this paper is how we generate the training inputs. The model needs both positive and negative contexts for the target keyword. They are called positive and negative samples. For producing the positive samples, we select combinations of annotated keywords from the document to which the target keyword belongs. In the case of the negative samples, we represent the keywords' co-occurrences in the dataset as a graph. In this graph, each keyword is a node and edges are created when two keywords appear together in a document. We select the negative samples from connected components different from that of the target keyword.

\begin{figure}
    \centering
    \includegraphics[width=\textwidth]{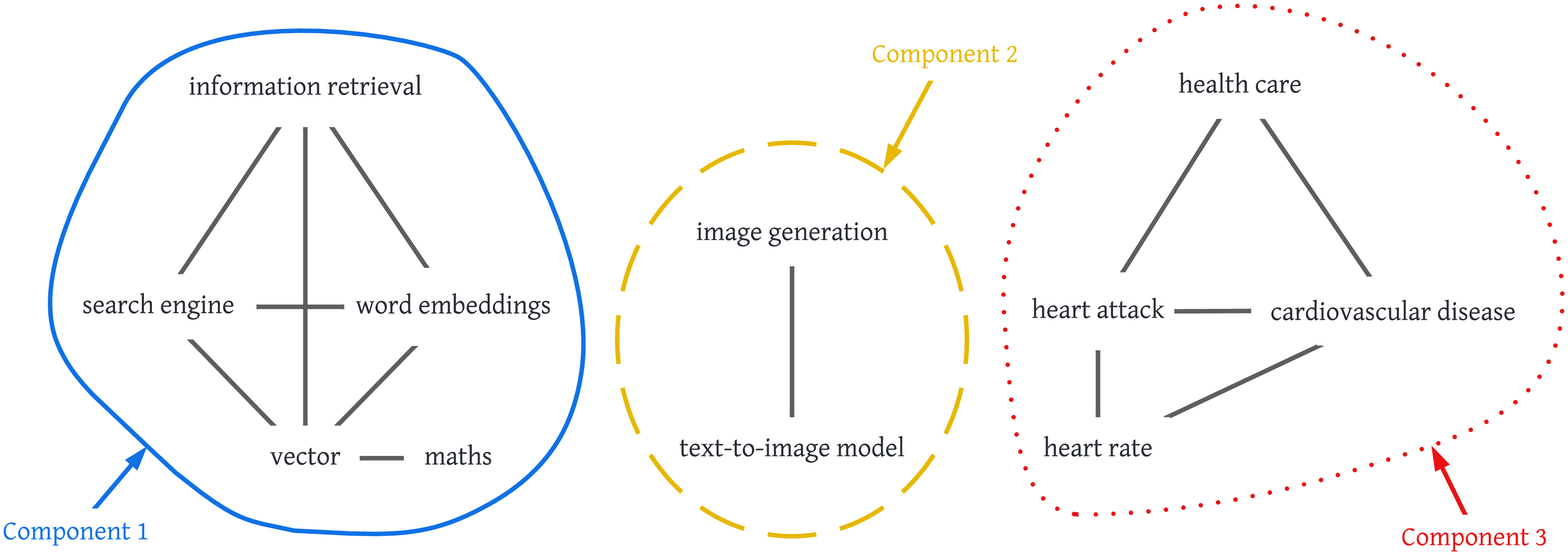}
    \caption{Example of a keyword co-occurrence graph with its connected components.}
    \label{fig:con_comps}
\end{figure}

Figure~\ref{fig:con_comps} shows an example of a keyword co-occurrence graph and its connected components. In that example, the positive samples for the keyword ``\textit{information retrieval}'' would be combinations of ``\textit{search engine}'', ``\textit{word embeddings}'' and ``\textit{vector}'', and its negative samples will always be extracted from connected components 2 or 3. Note that this means that the keyword ``\textit{maths}'' will never be a negative sample for ``\textit{information retrieval}'', even though it does not co-occur with it in any document.

\begin{figure}
    \centering
    \includegraphics[width=0.8\textwidth]{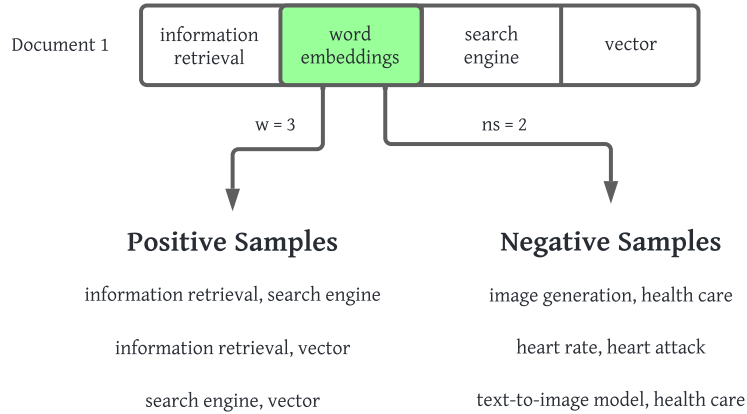}
    \caption{Positive and negative samples generation example. (\textit{The target keyword is highlighted}).}
    \label{fig:examples_gen}
\end{figure}

Figure~\ref{fig:examples_gen} shows how the positive and negative sampling is performed for a document extracted from the graph shown in the Figure~\ref{fig:con_comps}. Having the keyword ``\textit{word embeddings}'' as the target, we first select its positive samples. In this example, we select every combination of 2 ($w-1=2$) keywords that belong to the same document as the target. Then, for each pair of keywords we have to select the two negative samples ($ns=2$). Given the connected components shown in the Figure~\ref{fig:con_comps}, we select two keywords that do not belong to the same connected component than the target keyword.

Formally, to train these models, we use a dataset of scientific documents ($D$). Each document ($d \in D$) contains a set of annotated keywords ($K_d$). Then, given a combination size ($w-1$),  which plays an analogous role to the window size in the original models, we compute the set of positive samples ($K_d^{ps}$) of a document $d$ as follows:

\begin{equation*}
     K_d^{ps} = \Bigg\{(k_i,K_d^{ps_j}) \;\vert\; k_i \in K_d, \; K_d^{ps_j} \in {K_d - \{k_i\} \choose w - 1} \Bigg\},
\end{equation*}

\noindent that is, for each keyword ($k_i$) of the document's keywords set ($K_d$), we compute its positive samples ($K_d^{ps}$), which are the combinations of size $w - 1$ of the document's keywords set excluding the target keyword ($K_d - \{k_i\}$). Finally, for each positive samples set ($K_d^{ps_j}$) we obtain a pair $(k_i,K_d^{ps_j})$. 

As for the negative samples for each document ($K_d^{ns}$) we have followed the subsequent novel approach. First, we build the aforementioned keywords co-occurrence graph for the collection and compute its connected components. Then, for each pair $(k_i,K_d^{ps_j})$ we select as negative samples $ns$ keywords belonging to a different connected component than the target keyword ($k_i$).

To adapt the previous process to our \texttt{FastKeywords} model, we have to generate \textit{n}-grams for each context keyword (positive samples and negative samples) following the strategy explained before.

The \texttt{FastKeywords} model has two main advantages over \texttt{Keywords2Vec} because of the use of subword information:
\begin{itemize}
    \item It will perform significantly better on large collections.
    \item It will be able to generate embeddings for keywords that are not in the training corpus.
\end{itemize}

\begin{figure}
    \centering
    \includegraphics[width=\textwidth]{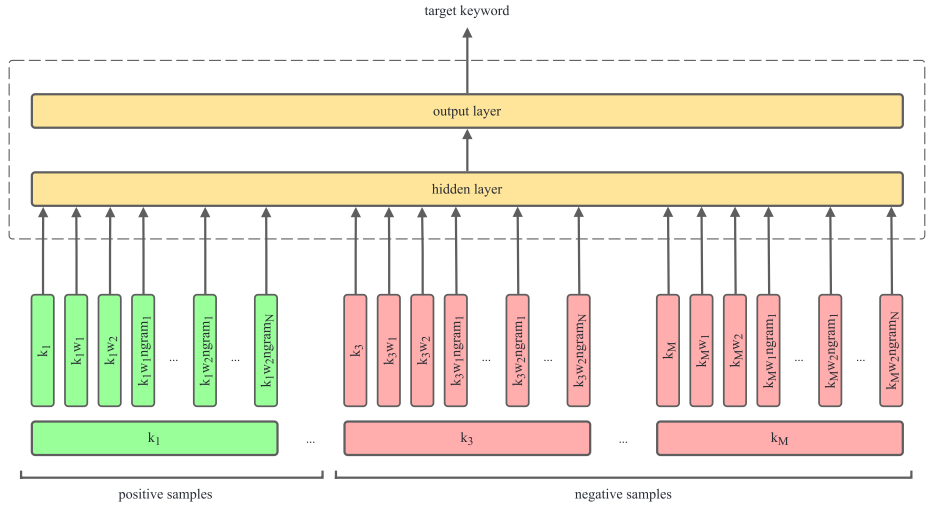}
    \caption{\texttt{FastKeywords} model high-level architecture diagram.}
    \label{fig:fastkwds_arch}
\end{figure}

Figure~\ref{fig:fastkwds_arch} shows the \texttt{FastKeywords} model's architecture. As we may see, it follows the classic CBOW strategy where several context samples are fed to the model in order to predict the target keyword.

\section{Experimental Setup}
This section describes the datasets used during the training and evaluation of the models, the baselines used to compare our model, the evaluation process and metrics, and finally, the parameters and setup used to train our models.

Our objective is to suggest similar keywords to the user input in the search process. With this in mind, we designed two experiments that approach the keyword suggestion problem as a keyword ranking task. In particular, the evaluation considers if the model can find keywords that belong to the same document as the target keyword. The rationale for this evaluation is that keywords with which the authors annotate a document tend to be semantically related. Alternatively, we may perform user studies to evaluate the perceived quality of the suggestions. We leave that for future work. 

\subsection{Datasets}
We selected two datasets commonly used on the classical keyword extraction task: the Inspec~\cite{hulth2003improved} and the KP20k~\cite{meng2017deep} collections.

\subsubsection{Inspec}
This collection consists of 2,000 titles and abstracts from scientific journal papers. The annotators assigned two sets of keywords to each document in this collection: controlled keywords that occur in the Inspec thesaurus and uncontrolled keywords that can be any suitable terms. We will only use the uncontrolled keywords created by professional indexers in our experiments, following the standard approach in the keyword extraction task. 

\subsubsection{KP20k}
This dataset is a keyword generation collection that includes 567,830 articles of the computer science domain obtained from various online digital libraries, including ACM Digital Library, ScienceDirect, Wiley and Web of Science. In this case, the authors made no differentiation during the annotation process, so we used all of the keywords in both training and testing.

\subsection{Baselines}
We include five baselines in our evaluation to compare our embedding models against other models that have already proven their effectiveness in capturing words or sentence syntax and semantic information.

The first baseline is the \texttt{Word2Vec} model trained in the Google News dataset and further tuned on each dataset to learn their vocabulary. That process was performed using the keywords set as a sentence so that the model could learn the contextual relationships between them.

The second baseline is the pre-trained \texttt{SBERT} \textit{all-mpnet-base-v2}\footnote{\url{https://huggingface.co/sentence-transformers/all-mpnet-base-v2}} model fine-tuned in the evaluation datasets for the sentence similarity task. To fine-tune the model, we served as inputs pairs of keywords and a binary value of similarity (1 if keywords belong to the same document and 0 if they do not).

The other three baselines were not trained for the task. These baselines are:
\begin{itemize}
    \item The pre-trained \texttt{SBERT} model, \textit{all-mpnet-base-v2}.
    \item The pre-trained \texttt{SciBERT} model that uses the uncased version of the vocabulary.
    \item The English \texttt{FastText} model trained on Common Crawl\footnote{\url{https://commoncrawl.org/}} and Wikipedia\footnote{\url{https://dumps.wikimedia.org/}}.
\end{itemize}

\subsection{Evaluation} \label{sec:eval}

As we mentioned before, to assess our models' performance in the keyword suggestion task, we devise two evaluation strategies representing two specific cases in the academic search field.

The first task consists of retrieving all the keywords with which a document was annotated, given one of them. The keyword retrieval is done via cosine similarity between the query keyword and all the other keywords. Then, we compute each average precision and aggregate all the results to compute the MAP (Mean Average Precision). Specifically, we calculate MAP@20.

The second task we propose follows a masking problem. For each document, we mask one keyword from the keywords set and then use the remaining keywords to try to find the masked one. For each of the remaining keywords, we retrieve the most similar ones using cosine similarity. Then, combining the scores of all the non-masked keywords, we compute the final ranking. Finally, we compute each masked keyword's reciprocal rank to calculate the MRR (Mean Reciprocal Rank). Precisely, we compute MRR@100. Note that for this evaluation method, we only used 50 random documents from the test split of each dataset.

Regarding methodologies, we evaluated both tasks under all-items and test-items approaches. In the all-items fashion, the keywords to be ranked are the whole set of keywords in the dataset, while in the test-items indexing approach, only the keywords in the test subset are ranked. 

We also perform statistical significance tests for all the evaluations. On the one hand, for comparing more than two systems, we use the Randomised Tukey HSD Test~\cite{tukey,tukey_book}. On the other hand, we follow the classic permutation test approach when comparing a pair of systems. These tests verify that the family-wise error does not exceed the confidence level $\alpha$.

\subsection{Experimental Settings}
This section will describe the parameters used to train the baselines and our proposed models. Table~\ref{table:training_parameters}  shows a summary of the models' training parameters detailed hereunder.

\begin{table*}[tb]
\centering
\caption{Models' training parameters on the Inspec and KP20k datasets.}
\label{table:training_parameters}
\setlength{\tabcolsep}{4pt}
\renewcommand{\arraystretch}{1.5}
\scriptsize
\begin{tabular}
{lccccc|ccccc}
\toprule
\multirow{2}*{Model} & \multicolumn{5}{c}{Inspec} & \multicolumn{5}{c}{KP20k}\\
 & \multicolumn{1}{|c}{Batch} & Epochs & Dim. & $w$ & $ns$  & Batch & Epochs & Dim. & $w$ & $ns$ \\
\midrule
\multicolumn{1}{l|}{\texttt{Word2Vec}}        & - & \textit{early stopping} & 300 & 6 & -  & - & 20 & 300 & 6 & -\\
\multicolumn{1}{l|}{\texttt{SBERT}}           & $2^{9}$ & \textit{early stopping} & 768 & - & 4 & $2^{9}$ & 20 & 768 & - & 4\\
\multicolumn{1}{l|}{\texttt{Keywords2Vec}}    & $2^{17}$ & \textit{early stopping} & 300 & 3 & 4 & $2^{17}$ & 20 & 300 & 3 & 4\\
\multicolumn{1}{l|}{\texttt{FastKeywords}}    & $2^{15}$ & \textit{early stopping} & 300 & 3 & 4 & $2^{15}$ & 20 & 300 & 3 & 4\\
\bottomrule
\end{tabular}
\end{table*}

To train the \texttt{Word2Vec} model, we followed an early-stopping strategy for the Inspec dataset, while for KP20k, we trained the model for 20 epochs. For both datasets, the selected embedding size was 300. Finally, we used a window size of 6 and 1 as the minimum count (lowest frequency of a word to be considered).

On the other hand, as we mentioned before, we fine-tuned the \texttt{SBERT} model in the sentence similarity task. For each positive sample (two keywords that co-occur in a document), we selected four negative samples (two keywords that never co-occur in a document). Again, we followed an early-stopping strategy for the Inspec dataset, while for KP20k, we trained the model for 20 epochs. For both datasets, we used a batch size of $2^{9}$ and an embedding size of 768.

We trained both our models using an early-stopping strategy for the Inspec dataset. On the other hand, when using the KP20k dataset, we trained each model for 20 epochs. Regarding the batch size, the \texttt{Keywords2Vec} model was trained with a batch size of $2^{17}$ on both datasets, while for the \texttt{FastKeywords} model, we used a batch size of $2^{15}$ on Inspec and KP20k. Also, both were trained to generate embeddings of size 300. The combination size and the number of selected negative samples were the same for both models on both datasets, 2 ($w=3$) and 4, respectively. Finally, regarding \texttt{FastKeywords} parameters, we used a minimum \textit{n}-gram size of 3, a maximum \textit{n}-gram size of 6 and a maximum number of \textit{n}-grams of 20.

When trained, we used the whole set of document keywords from the dataset in the tuning process of the models\footnote{All experiments were run on an NVIDIA A100 GPU with 80GB of memory.}. When not trained, we used the default pre-trained models for the baselines. In the evaluation process, we selected a subset of documents on which we performed the aforementioned tasks.

\section{Results}
This section reports how \texttt{Keywords2Vec} and \texttt{FastKeywords} perform against the selected baselines and how they perform against each other. As we mentioned in Section~\ref{sec:eval} we report results using two tasks and two evaluation techniques.

Table~\ref{table:map_results} shows the results for the document's keywords identification task. We can see that our models significantly outperform the established baselines in both datasets and evaluation approaches. This is especially remarkable in the case of the state-of-the-art sentence BERT model, which, even for the fine-tuned scenario and using embeddings with twice the dimensions, lies quite behind our proposals. As expected, using the test-items strategy produces better results since the keyword set where the search is performed is much smaller.

Moreover, we can see that the dataset size is a determinant factor in this task. A much larger set of keywords (this is the case of the KP20k dataset versus the Inspec dataset) greatly impacts the final score. This makes sense because increasing the keyword set size makes finding the ones we are looking for more challenging (something analogous happens to a lesser degree between the all-items and test-items results).

In terms of comparing \texttt{FastKeywords} against \texttt{Keywords2Vec} we can see that the former performs better and that the more considerable difference between them appears when the keyword set is the biggest (KP20k dataset). The main reason behind this relies on the capability of \texttt{FastKeywords} to leverage subword information to build the keywords embeddings.
\begin{table*}[tb]
\centering
\caption{MAP@20 for the document's keywords identification task. Statistically significant improvements according to the Randomized Tukey HSD test ($permutations = 100,000$; $\alpha = 0.05$) are superscripted.}
\label{table:map_results}
\setlength{\tabcolsep}{5pt}
\renewcommand{\arraystretch}{1.5}
\scriptsize
\begin{tabular}
{lll|ll}
\toprule
\multirow{2}*{Model} & \multicolumn{2}{c} {Inspec} & \multicolumn{2}{c}{KP20K} \\
 & \multicolumn{1}{|c} {All items} & \multicolumn{1}{c|} {Test items} &  \multicolumn{1}{c} {All items} & \multicolumn{1}{c} {Test items} \\
\midrule
\multicolumn{1}{l|} {\texttt{SBERT} (\textit{no train}) ($\dagger$)}       & 0.0323 & $0.0544^\uplus$ & 0.0039 & $0.0134^\uplus$ \\
\multicolumn{1}{l|} {\texttt{SciBERT} (\textit{no train}) ($\ddagger$)}    & 0.0205 & 0.0345 & 0.0040 & 0.0110 \\
\multicolumn{1}{l|} {\texttt{FastText} (\textit{no train}) ($\uplus$)}     & 0.0147 & 0.0217 & 0.0039 & 0.0095 \\
\multicolumn{1}{l|} {\texttt{Word2Vec} ($\mp$)}                            & $0.0463^\uplus$ & $0.0832^{\dagger\ddagger\uplus}$ & 0.0048 & $0.0130^{\ddagger\uplus}$ \\
\multicolumn{1}{l|} {\texttt{SBERT} ($\pm$)}                               & $0.1481^{\dagger\ddagger\uplus\mp}$ & $0.5880^{\dagger\ddagger\uplus\mp}$ & $0.0072^{\dagger\ddagger\uplus\mp}$ & $0.0182^{\dagger\ddagger\uplus\mp}$ \\ \midrule
\multicolumn{1}{l|} {\texttt{Keywords2Vec} ($\otimes$)}           & $0.8634^{\dagger\ddagger\uplus\mp\pm}$ & $0.9090^{\dagger\ddagger\uplus\mp\pm}$ & $0.0690^{\dagger\ddagger\uplus\mp\pm}$ & $0.0918^{\dagger\ddagger\uplus\mp\pm}$ \\
\multicolumn{1}{l|} {\texttt{FastKeywords} ($\odot$)}            & $\textBF{0.8659}^{\dagger\ddagger\uplus\mp\pm}$ & $\textBF{0.9161}^{\dagger\ddagger\uplus\mp\pm}$ & $\textBF{0.0762}^{\dagger\ddagger\uplus\mp\pm\otimes}$ & $\textBF{0.1060}^{\dagger\ddagger\uplus\mp\pm\otimes}$ \\
\bottomrule
\end{tabular}
\end{table*}

Table~\ref{table:mrr_results} shows results for the masked keyword discovering task. These results confirm what we had already seen in the first one: our models significantly outperform all the established baselines in every experiment we performed. Again, we can see that increasing the keyword set length produces worse results as the task becomes more and more challenging. Again, when comparing the proposed models against each other on this task, we can see that \texttt{FastKeywords} performs better than \texttt{Keywords2Vec} when increasing the dataset size, getting statistically significant improvements over the \texttt{Word2Vec}-based method.

\begin{table*}[tb]
\centering
\caption{MRR@100 for the masked keyword discovering task. Statistically significant improvements according to the Randomized Tukey HSD test ($permutations = 1,000,000$; $\alpha = 0.05$) are superscripted.}
\label{table:mrr_results}
\setlength{\tabcolsep}{5pt}
\renewcommand{\arraystretch}{1.5}
\scriptsize
\begin{tabular}
{lll|ll}
\toprule
\multirow{2}*{Model} & \multicolumn{2}{c} {Inspec} & \multicolumn{2}{c}{KP20K} \\
 & \multicolumn{1}{|c} {All items} & \multicolumn{1}{c|} {Test items} &  \multicolumn{1}{c} {All items} & \multicolumn{1}{c} {Test items} \\
\midrule
\multicolumn{1}{l|} {\texttt{SBERT} \textit{no train} ($\dagger$)}     & 0.0342 & 0.0479 & 0.0018 & 0.0152 \\
\multicolumn{1}{l|} {\texttt{SciBERT} \textit{no train} ($\ddagger$)}  & 0.0220 & 0.0345 & 0.0056 & 0.0129 \\
\multicolumn{1}{l|} {\texttt{FastText} \textit{no train} ($\uplus$)}   & 0.0068 & 0.0097 & 0.0023 & 0.0047 \\
\multicolumn{1}{l|} {\texttt{Word2Vec} ($\mp$)}                        & 0.0751 & $0.1187^\uplus$ & 0.0008 & 0.0079 \\
\multicolumn{1}{l|} {\texttt{SBERT} ($\pm$)}                           & $0.1688^{\dagger\ddagger\uplus}$ & $0.5890^{\dagger\ddagger\uplus\mp}$ & 0.0138 & 0.0448 \\ \midrule
\multicolumn{1}{l|} {\texttt{Keywords2Vec} ($\otimes$)}       & $0.8914^{\dagger\ddagger\uplus\mp\pm}$ & $0.9100^{\dagger\ddagger\uplus\mp\pm}$ & $0.0778^{\dagger\ddagger\uplus\mp\pm}$ & $0.0828^{\dagger\ddagger\uplus\mp}$ \\
\multicolumn{1}{l|} {\texttt{FastKeywords} ($\odot$)}         & $\textBF{0.8988}^{\dagger\ddagger\uplus\mp\pm}$ & $\textBF{0.9102}^{\dagger\ddagger\uplus\mp\pm}$ & $\textBF{0.1402}^{\dagger\ddagger\uplus\mp\pm\otimes}$ & $\textBF{0.1467}^{\dagger\ddagger\uplus\mp\pm\otimes}$ \\
\bottomrule
\end{tabular}
\end{table*}

Table~\ref{table:ns_results} shows the performance of the proposed negative sampling method, comparing it with a \texttt{FastKeywords} model that uses a random negative sampling strategy. The results ratify that the proposed method works significantly better than a naive random approach in all cases except on MRR@100 using the all-items indexing strategy. For this case, the \textit{p}-value is 0.06.

\begin{table*}[tb]
\centering
\caption{Negative sampling methods comparison on the Inspec dataset. Statistically significant improvements according to the permutation test ($permutations = 1,000,000$; $\alpha = 0.05$) are superscripted.}
\label{table:ns_results}
\setlength{\tabcolsep}{5pt}
\renewcommand{\arraystretch}{1.5}
\scriptsize
\begin{tabular}
{lll|ll}
\toprule
\multirow{2}*{Model} & \multicolumn{2}{c} {MAP@20} & \multicolumn{2}{c}{MRR@100} \\
 & \multicolumn{1}{|c} {All items} & \multicolumn{1}{c|} {Test items} &  \multicolumn{1}{c} {All items} & \multicolumn{1}{c} {Test items} \\
\midrule
\multicolumn{1}{l|} {\texttt{FastKeywords} \textit{random negative sampling} ($\oplus$)}   & 0.8420 & 0.9093 & 0.8750 & 0.8890 \\
\multicolumn{1}{l|} {\texttt{FastKeywords} ($\odot$)}             & $\textBF{0.8659}^\oplus$ & $\textBF{0.9161}^\oplus$ & \textBF{0.8988} & $\textBF{0.9102}^\oplus$ \\
\bottomrule
\end{tabular}
\end{table*}

Finally,  for illustrative purposes, Table~\ref{table:nn} shows the top 10 nearest neighbours for the keyword ``\textit{information retrieval}'' on the KP20k collection retrieved by both proposed models. We also show each keyword's associated score, which represents the similarity with the query keyword.

\begin{table*}[tb]
\centering
\caption{Top 10 nearest neighbours for the keyword ``\textit{information retrieval}'' on the KP20k collection (\textit{test-items index}).}
\label{table:nn}
\renewcommand{\arraystretch}{1.5}
\scriptsize
\begin{tabularx}{\textwidth}
{>{\centering\arraybackslash}m{0.25\textwidth}>{\centering\arraybackslash}m{0.1\textwidth}p{0.1\textwidth}>{\centering\arraybackslash}m{0.4\textwidth}>{\centering\arraybackslash}m{0.1\textwidth}}
\multicolumn{2}{c}{\texttt{FastKeywords}} & & \multicolumn{2}{c}{\texttt{Keywords2Vec}} \\
\cline{1-2} \cline{4-5}
\textbf{keyword} & \textbf{score} & & \textbf{keyword} & \textbf{score} \\
\cline{1-2} \cline{4-5}
ranking & 0.9133 & &                  retrieval model & 0.6045 \\
query expansion & 0.9104 & &         collection selection & 0.5587 \\
text classification & 0.9067 & &     link topic detection & 0.5472\\
relevance & 0.9044 & &                dempstershafer evidence theory & 0.5418 \\
text mining & 0.9018 & &             retrospective evidence event detection & 0.5371 \\
information extraction & 0.9012 & &  instance based learning & 0.5361 \\
relevance feedback & 0.8980 & &      query expansion & 0.5282 \\
knowledge discovery & 0.8979 & &     terminology extraction & 0.5260 \\
document clustering & 0.8971 & &     viral marketing & 0.5253 \\
text categorization & 0.8970 & &     query formulation & 0.5248 \\
\cline{1-2} \cline{4-5}
\end{tabularx}
\end{table*}

\section{Conclusions}
This paper explored the potential of keyword embedding models in the keyword suggestion task. We also propose several baselines and new evaluation methods to assess the performance of the models, as not much previous work has been published for this task.

The proposed models adapt \texttt{Word2Vec} and \texttt{FastText} CBOW architectures to compute keyword embeddings instead  of word embeddings. Along with these two keyword embedding models, we present a novel strategy for the negative sampling task, which leverages the potential of the keyword co-occurrence graph's connected components to perform a better selection of the negative samples.

Results show that our methods significantly outperform the selected baselines on both evaluation datasets. We also demonstrated the potential of sub-keyword and sub-word information to represent keywords as embeddings. In future work we aim to:
\begin{itemize}
    \item Use the designed weights system to give more relevance to full keywords and words than to \textit{n}-grams.
    \item Assess the models' performance using popularity-based negative sampling.
    \item Combine negative samples extracted from the target keyword connected component and from different connected components.
    \item Use special delimiters to differentiate if a word is a part of a keyword or a keyword itself or if a \textit{n}-gram is a part of a word or a word itself.
    \item Train and test the models on non-scientific keyword-style annotated data.
    \item Study how the offline findings of this work align with live user testing.
\end{itemize}

\subsubsection{Acknowledgements} This work was supported by projects PLEC2021-007662 (MCIN/AEI/10.13039/501100011033, Ministerio de Ciencia e Innovación, Agencia Estatal de Investigación, Plan de Recuperación, Transformación y Resiliencia, Unión Europea-Next Generation EU) and RTI2018-093336-B-C22 (Ministerio de Ciencia e Innovación, Agencia Estatal de Investigación). The first and third authors also thank the financial support supplied by the Consellería de Cultura, Educación e Universidade Consellería de Cultura, Educación, Formación Profesional e Universidades (accreditation 2019-2022 ED431G/01, ED431B 2022/33) and the European Regional Development Fund, which acknowledges the CITIC Research Center in ICT of the University of A Coruña as a Research Center of the Galician University System. The fist author also acknowledges the support of grant DIN2020-011582 financed by the MCIN/AEI/10.13039/501100011033.

%
%
%

\end{document}